\documentclass{PoS}

\newcommand*{\chipt}{{$\chi$PT }}

\def\etal{{\it et al.}}

\def\CO{{\cal O}}

\def\bar{\overline}

\def\spose#1{\hbox to 0pt{#1\hss}}
\def\ltapprox{\mathrel{\spose{\lower 3pt\hbox{$\mathchar''218$}}
 \raise 2.0pt\hbox{$\mathchar''13C$}}}
\def\gtapprox{\mathrel{\spose{\lower 3pt\hbox{$\mathchar''218$}}
 \raise 2.0pt\hbox{$\mathchar''13E$}}}
\def\inapprox{\mathrel{\spose{\lower 3pt\hbox{$\mathchar''218$}}
 \raise 2.0pt\hbox{$\mathchar''232$}}}

\newcommand{\bra}[1]{\langle #1|}
\newcommand{\ket}[1]{|#1\rangle}

\title{Form factors for $B$ to $Kll$ semileptonic decay from three-flavor lattice QCD}

\ShortTitle{Form factors for B to Kll semileptonic decay from three-flavor lattice QCD}

\author{\speaker{Ran Zhou}\\
  Physics Department, Indiana University\\
        E-mail: \email{zhouran@indiana.edu}}

\author{Jon A. Bailey \\  Physics Department, Seoul National University}
\author{Alexei Bazavov\\   Physics Department, Brookhaven National Laboratory}
\author{Aida X. El-Khadra\\  Physics Department, University of Illinois}
\author{Steven Gottlieb\\  Physics Department, Indiana University}
\author{Rajendra D. Jain\\    Physics Department, University of Illinois}
\author{Andreas S. Kronfeld\\   Theoretical Physics Department, Fermi National Accelerator Laboratory}
\author{Ruth S. Van de Water\\  Physics Department, Brookhaven National Laboratory}
\author{(Fermilab Lattice and MILC Collaborations)}

\abstract{We study the $B \to Kl^+l^-$ semileptonic decay process in three-flavor
lattice QCD. We analyze several ensembles generated by the MILC
collaboration at different lattice spacings and sea-quark masses.
We use the  asqtad improved staggered action for the light quarks
and the clover action with the Fermilab interpretation 
for the heavy $b$ quark. We present preliminary results for the 
vector current induced form factors for a range of kaon energies. Our analysis
includes chiral and continuum extrapolations based on SU(2) staggered \chipt.}

\FullConference{ The XXIX International Symposium on Lattice Field Theory - Lattice 2011\\
July 10-16, 2011\\
Squaw Valley, Lake Tahoe, California}

\begin{document}

\section{Introduction}
Rare decays of $B$ or $D$ mesons may play an important role in the discovery 
of new physics in the flavor sector. 
Transitions in which a $b$ quark decays to an $s$ quark proceed
through a flavor changing neutral current.  As these currents
only occur at higher order within the Standard Model (SM), such
decays are rare, and a small contribution from new physics beyond the
Standard Model may be comparable to the SM contribution and
hence observable. 
In this work, we focus on the $B\to K l^+l^-$ semileptonic decay process, 
which occurs via the transition of $b\to sl^+l^-$ at the quark level. 
References~\cite{Hurth:2010tk} and~\cite{Antonelli:2009ws} summarize both the
experimental and theoretical status of $B$-meson decays.
The BABAR Collaboration studied both $B \to Kl^+l^-$ and $B \to K^*l^+l^-$ 
semileptonic decays in Ref.~\cite{Aubert:2008ps}. The Belle Collaboration 
published their results on the same process in Ref.~\cite{:2009zv}.
The CDF Collaboration studied the $B\to K^*\mu^+\mu^-$ decay
in Ref.~\cite{Aaltonen:2011cn}. 

The $B\to Kl^+l^-$ form factors are calculable from first principles
using lattice QCD as there is only one hadron in the initial state and one 
in the final state. 
Recently, calculations using the MILC 2+1 flavor dynamical quark
ensembles have been done for $B \to K^*l^+l^-$
by Liu \etal ~\cite{Liu:2011ra} and a study of $B\to K^* \gamma$ form factors 
was done by Becirevic \etal ~\cite{Becirevic:2006nm}.
The Fermilab Lattice and MILC Collaborations presented some 
preliminary results for the $B\to Kl^+l^-$ form factors 
in Ref.~\cite{Jain:2006zz}. 
Additional ensembles covering a wider range of lattice
spacings have been analyzed since then. 
In this brief report, we show a more comprehensive (but still
preliminary) analysis of the meson masses, form factors and chiral
and continuum extrapolations. 

\section{Theoretical Background}
An operator production expansion (OPE) analysis of $B\to Kl^+l^-$ 
shows that two currents, a vector current
$\bar{s}\gamma^\mu b$ and a
tensor current $\bar{s} \sigma^{\mu\nu}q_\nu b$ 
contribute to this process at lowest order~\cite{Hurth:2010tk}. 
We study the vector current here and defer study of the
tensor current to later work.
The matrix element of the vector current can be expressed in terms of
two form factors $f_+$ and $f_0$ as:
\begin{eqnarray}
\bra{K} i\bar{s}\gamma^\mu b \ket{B}&=f_+(q^2)\left(p_B^\mu+p_K^\mu-\frac{m_B^2-m_K^2}{q^2}q^\mu \right)
+f_0(q^2)\frac{m_B^2-m_K^2}{q^2}q^\mu,
\end{eqnarray}
where $q^\mu=p_B^\mu-p_K^\mu$. 
We study the form factors in the the $B$-meson rest frame,
so only the kaon has non-zero momentum. 
The form factors $f_\parallel$ and $f_\perp$ are defined as:
\begin{eqnarray}
f_\parallel&=\frac{\bra{K} i\bar{s} \gamma^0 b \ket{B}}{\sqrt{2m_B}} ,\\ 
f_\perp&=\frac{\bra{K} i\bar{s}\gamma^i b \ket{B}}{\sqrt{2m_B}p^i_K} .
\end{eqnarray}
They are more convenient for our lattice calculation, and
are related to $f_+$ and $f_0$ by:
\begin{eqnarray}
f_+ & = & \frac{1}{\sqrt{2m_B}}\left [ f_\parallel+(m_B-E_K)f_\perp \right] ,\\
f_0 & = & \frac{\sqrt{2m_B}}{m_B^2-m_K^2}\left [ (m_B-E_K)f_\parallel+(E_K^2-m_K^2)f_\perp \right] .
\end{eqnarray}
The form factor $f_\perp$, as compared to $f_\parallel$, gives the dominant 
contribution to $f_+$, and hence to the experimental decay rate.

The lattice form factors $f_\parallel$ and $f_\perp$ are computed numerically 
at several values of the lattice spacing and of the average up-down and strange 
quark masses.  These results must then be extrapolated to the physical quark masses 
and the continuum using chiral perturbation theory ($\chi$PT). ~$SU(3)$ staggered 
chiral perturbation theory for heavy-light semileptonic form factors~\cite{Aubin:2007mc} was 
successfully applied to the case of $B\to\pi\ell\nu$ decay in Ref.~\cite{Bailey:2008wp}.  More 
recently, $SU(2)$ $\chi$PT was applied to the extrapolation of $D\to\pi\ell\nu$ 
form factors on $N_f=2$ lattices~\cite{DiVita:2011py}. Some studies purport that 
$SU(2)$ \chipt may be a better effective theory for heavy-light 
physics projects~\cite{Becirevic:2006me}. Here we test both $SU(3)$ and $SU(2)$ 
formalisms in our $B\to Kl^+l^-$ chiral-continuum extrapolations. 

\section{Numerical Simulation}
\begin{table}[t]
 \centering
 \begin{tabular}{ccccc}
  \hline  
  \hline  
  $a$~(fm) & $am_l^{\rm sea}/am_s^{\rm sea}$ & $am_l^{\rm val}/am_s^{\rm val}$ & $\kappa_b$ & $N_{\rm measure}$ \\
  \hline
  0.12 & 0.02/0.05 & 0.02/(0.0415, 0.05) & 0.0918 & 2052\\ 
  0.12 & 0.01/0.05 & 0.01/(0.0415, 0.05) & 0.0901 & 2259\\
  0.12 & 0.007/0.05 & 0.005/(0.0415, 0.05) & 0.0901 & 2110\\
  0.12 & 0.005/0.05 & 0.005/(0.0415, 0.05) & 0.0901 & 2099\\
  \hline
  0.09 & 0.0124/0.031 & 0.0124/(0.0261, 0.0310) & 0.0982 & 1996\\
  0.09 & 0.0062/0.031 & 0.0062/(0.0261, 0.0310) & 0.0979 & 1931\\
  0.09 & 0.00465/0.031 & 0.0047/(0.0261, 0.0310) & 0.0997 & 984\\
  0.09 & 0.0031/0.031 & 0.0031/(0.0261, 0.0310) & 0.0976 & 1015\\
  0.09 & 0.00155/0.031 & 0.00155/(0.0261, 0.0310) & 0.0976 & 791\\
  \hline
  0.06 & 0.0072/0.018 & (0.0072)/(0.0188) & 0.1048 & 593 \\
  0.06 & 0.0018/0.018 & (0.0018)/(0.0188)  & 0.1052 & 827\\
  \hline  
  \hline
 \end{tabular}
 \caption{Ensembles of QCD gauge field configurations used in the current $B\to Kl^+l^-$ work. 
   $am_l^{\rm sea}$ and $am_s^{\rm sea}$ denote the light and strange sea
quark masses.
   $am_l^{\rm val}$ and $am_s^{\rm val}$ denote spectator and daughter 
quark masses in the $b \to s$ transition. 
Both unitary and partially quenched kaon mass points are included in the 
chiral-continuum extrapolations. }
 \label{tab:ensembles}
\end{table}
Our lattice calculations are done on MILC's $N_f$=2+1 flavor 
gauge configurations with asqtad improved quarks~\cite{Bazavov:2009bb}. 
The clover action with the Fermilab interpretation 
is used for the heavy quark~\cite{ElKhadra:1996mp}. 
MILC's ensembles cover
many lattice spacings, light quark masses and volumes, which allows good
control on the form factors' chiral and continuum extrapolations. 
The $b$ quark mass  is tuned close to its physical value as in Ref.~\cite{Bernard:2010fr}. 
In this report, we include results from the coarse ($a\approx 0.12$~fm), 
fine ($a\approx 0.09$~fm), and superfine ($a\approx 0.06$~fm) ensembles.

\section{Numerical Results}
The first step of our analysis is to determine the $B$ meson 
masses and kaon masses and energies from fits to the two-point correlators on every ensemble.
States of both positive and negative parity contribute to
the two-point correlators~\cite{Bailey:2008wp}. We vary the number of 
states in the fits and select the fit range from $t_{\rm min}$ to $t_{\rm max}$ 
carefully to control the exited state contribution and obtain a 
good $p$-value (confidence level) of the fit. Finally, the error on the mass 
is estimated via the standard jackknife method.

The second step of the analysis is to extract the form factors 
$f_\parallel$ and $f_\perp$ from fits of the ratio of 
three-point and two-point functions. The three-point function is defined as:
\begin{eqnarray}
  C_{3,\mu}^{B\to K}(t,T; \vec{p}_K) & = & \sum_{\vec{x},\vec{y}} e^{i \vec{p}_K \cdot \vec{y}} \langle \CO_K (0,\vec{0})\, V_\mu (t,\vec{y})\, \CO^\dagger_B (T,\vec{x}) \rangle ,
\label{eq:R3mu}
\end{eqnarray}
where $V_\mu$=$i \bar{s} \gamma_\mu b$ and $T$ is the location of the sink
operator.  Because we study the form factors in $B$-meson rest frame, 
only the kaon has non-zero momentum ($p_K$). 
In a finite volume, the kaon's momentum is discrete. 
We choose $p=(0,0,0)2\pi/L$, $(1,0,0)2\pi/L$, and $(1,1,0)2\pi/L$, where $L$ is the box size.
Higher momentum data are omitted due to their large statistical fluctuations. 
In addition, an iterative averaging trick is used to suppress 
the contribution of the states that have an 
alternating sign~\cite{Bailey:2008wp}. The ratio of two and three-point 
functions is defined as:
\begin{eqnarray}
        \bar{R}_{3,\mu}^{B\to K} (t,T) & \equiv & \frac{\bar{C}_{3,\mu}^{B\to K} (t,T)}{\sqrt{\bar{C}_{2}^{K} (t)\bar{C}_{2}^{B} (T-t)}} \sqrt{\frac{2 E_K}{e^{-E_K^{(0)}t} \, e^{-m_B^{(0)}(T-t)}}} .
\end{eqnarray}
where $\bar{C}_2$ and $\bar{C}_3$ are averaged correlation functions~\cite{Bailey:2008wp}. 
After multiplying $\bar{R}$ by the required renormalization constant, 
we obtain the continuum form factors:\footnote{The calculation of the renormalization 
constants on the superfine ensembles has not been finished yet. We 
use the values from the coarse and fine ensembles to estimate 
these factors on the superfine ensembles. We also set $\rho$ as 
1 for this current preliminary analysis.}
\begin{eqnarray}
\label{eq:fpar_R}
        f_\parallel^\textrm{cont} & = &\rho\sqrt{Z_V^{hh}Z_V^{ll}}\bar{R}_{3,0}^{B\to K} (t,T) , \\
        f_\perp^\textrm{cont} & = &\rho \sqrt{Z_V^{hh}Z_V^{ll}}\frac{1}{p^i_\pi} \, \bar{R}_{3,i}^{B\to K} (t,T)  .
\label{eq:fperp_R}
\end{eqnarray}

\begin{figure}[ht]
\centering
\includegraphics[scale=0.85]{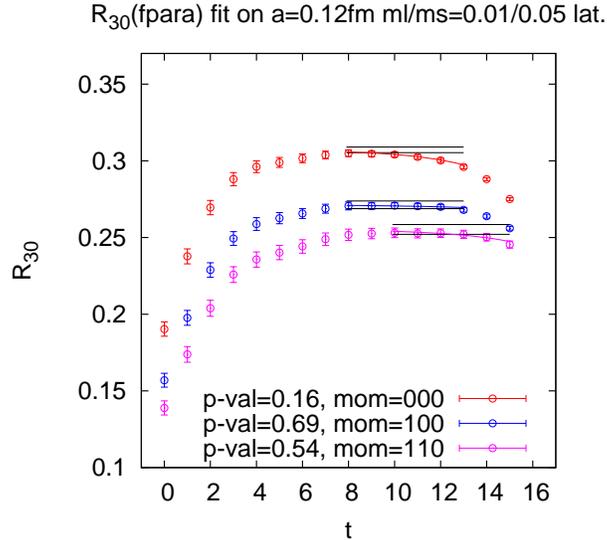}
\caption{Example $f_\parallel$ fit 
on the coarse ($a=0.12$~fm), $am_l/am_s$=0.01/0.05 ensemble. 
The $y$-axis is the ratio $R_{3,0}$ [c.f. Eq. ({\protect\ref{eq:R3mu}})] without 
any renormalization factors. }
\label{fig:fpara}
\end{figure}
Figure~\ref{fig:fpara} shows an example $f_\parallel$ fit on the
coarse, $am_l/am_s$=0.01/0.05 ensemble with 
$p=(0,0,0)2\pi/L$, $(1,0,0)2\pi/L$, and $(1,1,0)2\pi/L$. The $y$-axis is 
$R_{3,0}$ without multiplication by the renormalization  constants. 
We fit $R_{3,0}$ with a constant term plus an exponential decay term. 
We choose the preferred fit range by fixing the size the fit interval, 
{\it i.e.}, $t_{\rm max} - t_{\rm min}$, but shifting the location of the 
minimum time slice $t_{\rm min}$ to obtain a stable central value and errors and a good $p$-value.
We plot the the result of the constant term and its error, 
which is consistent with 
the single plateau fit method used in Ref.~\cite{Bailey:2008wp}. 

\begin{figure}[ht]
\begin{minipage}[b]{0.5\linewidth}
\centering
\includegraphics[scale=0.65]{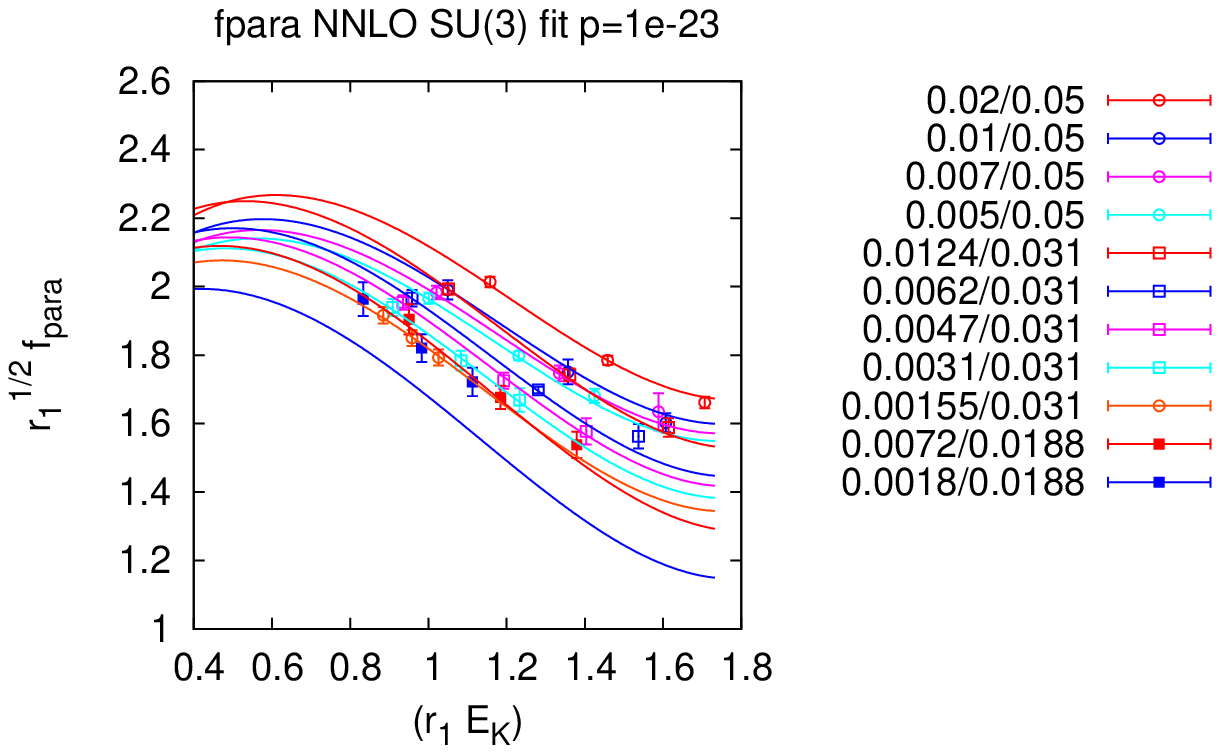}
\end{minipage}
\hspace{0.5cm}
\begin{minipage}[b]{0.5\linewidth}
\centering
\includegraphics[scale=0.65]{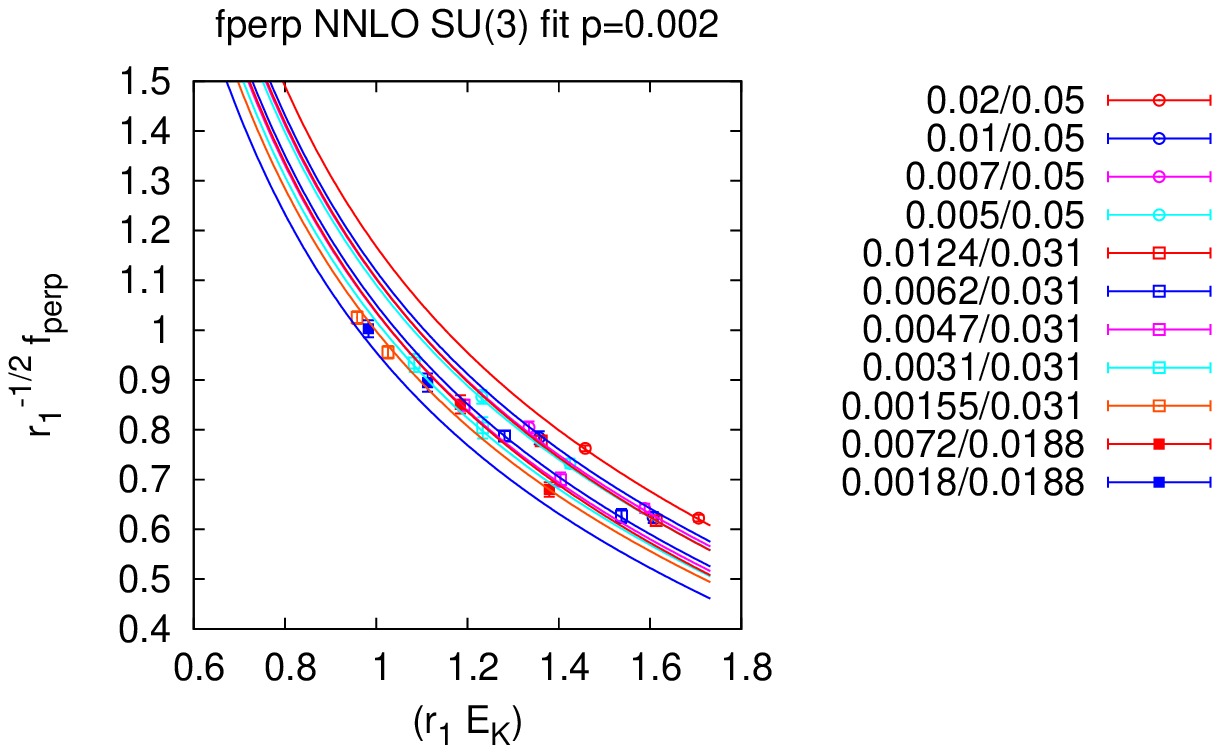}
\end{minipage}
\caption{$f_\parallel$ (left panel) and $f_\perp$ (right panel) chiral-continuum extrapolations with 
NNLO $SU(3)$ S$\chi$PT. Partially-quenched points are included in the fits, but are not shown in 
the figures for clarity.  Open circles denote coarse data points, open squares denote fine data, 
and filled squares denote superfine data.  Fit lines should pass through the data points of the 
corresponding color.}
\label{fig:su3_chpt_fit}
\end{figure}

\begin{figure}[ht]
\begin{minipage}[b]{0.5\linewidth}
\centering
\includegraphics[scale=0.65]{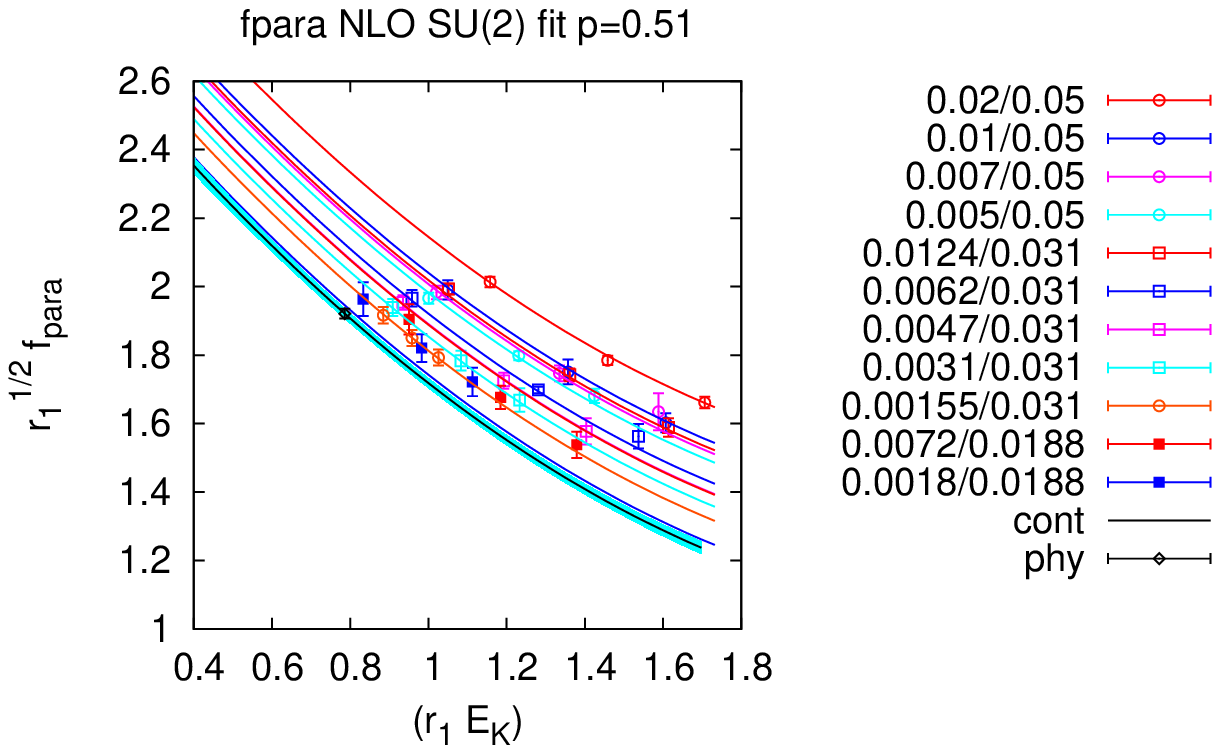}
\end{minipage}
\hspace{0.5cm}
\begin{minipage}[b]{0.5\linewidth}
\centering
\includegraphics[scale=0.65]{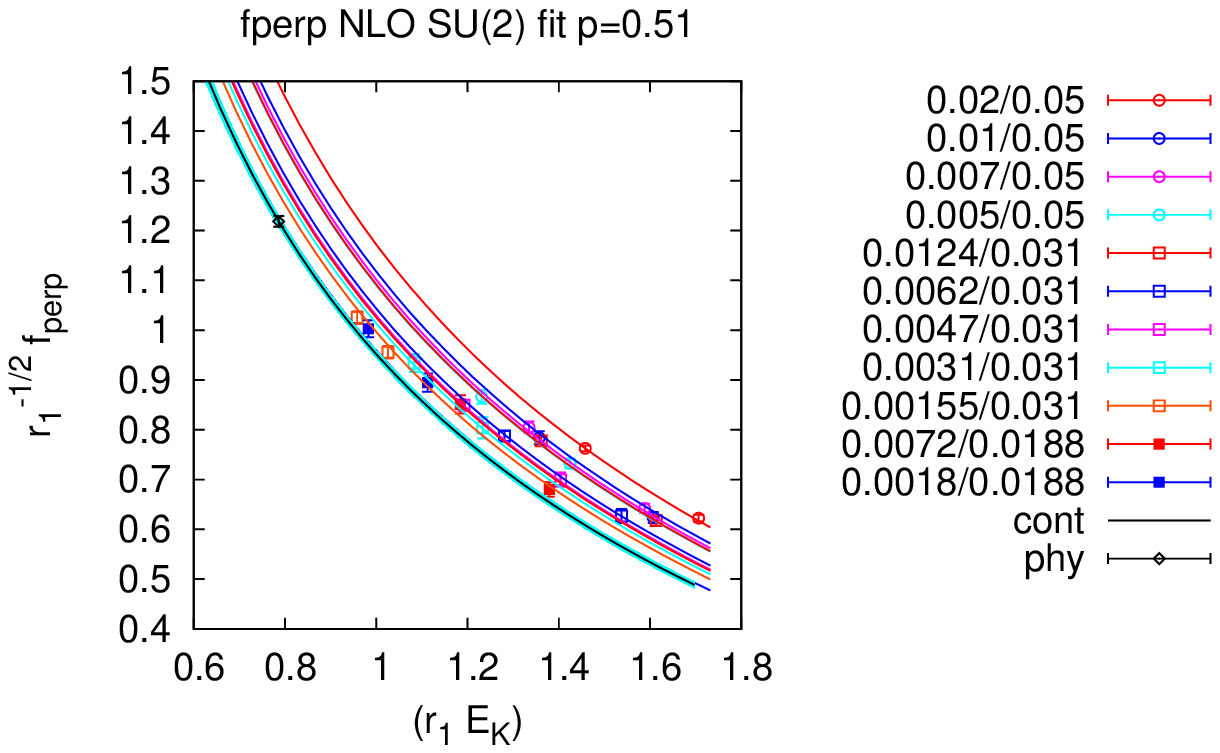}
\end{minipage}
\caption{$f_\parallel$ (left panel) and $f_\perp$ (right panel) chiral-continuum extrapolations with 
NLO $SU(2)$ S$\chi$PT. Partially-quenched points are included in the fits, but are not shown in 
the figures for clarity.  Open circles denote coarse data points, open squares denote fine data, 
and filled squares denote superfine data.  Fit lines should pass through the data points of the 
corresponding color.  The cyan band shows the continuum-extrapolated form factor at the 
physical light-quark mass with statistical errors.}
\label{fig:chpt_fit}
\end{figure}

To determine the form factors defined in the continuum with physical 
quark masses, a combined chiral and continuum extrapolations is applied. 
We use Staggered Chiral Perturbation Theory (S\chipt) as the low energy 
effective theory. S\chipt accounts for taste symmetry breaking effects 
in the staggered quark action. NLO $SU(3)$ S\chipt supplemented by 
NNLO analytic terms was used
successfully in $B\to \pi$ and $D\to \pi$ 
decays~\cite{Bailey:2008wp, Bailey:2010vz}, but we find that it fails in the 
$f_\parallel$ fit for $B\to Kl^+l^-$ process. Figure~\ref{fig:su3_chpt_fit}
shows the of result of the NNLO $SU(3)$ S\chipt fit for $f_\parallel$. 
The $SU(3)$ fit gives a low $p$-value and incorrect
behavior in the small $E_K$ region, which may indicate the
fact that the kaon is too heavy for $SU(3)$ S\chipt. On the other hand, 
$SU(3)$ \chipt gives a reasonable extrapolation for $f_\perp$, since 
the shape of $f_\perp$ is dominated by the $B^*$ pole and $SU(3)$ S\chipt accounts for it 
correctly. However, the $p$-value is still poor. 

Given these difficulties with $SU(3)$ S$\chi$PT, we switch to $SU(2)$ 
S\chipt for the chiral-continuum extrapolations.
Our $SU(2)$ formula is inspired by the work of Ref.~\cite{DiVita:2011py}. We 
take the heavy $m_s$ limit of the $SU(3)$ S\chipt result~\cite{Aubin:2007mc} to obtain the SU(2) 
reduction. Figure \ref{fig:chpt_fit} shows that $SU(2)$ S\chipt works well with our data even at NLO 
and provides  better control in the low $E_K$ region. The results here are still 
preliminary. The detailed comparisons between $SU(3)$ and $SU(2)$ fits are still 
under investigation.

\begin{figure}[ht]
\centering
\includegraphics[scale=0.85]{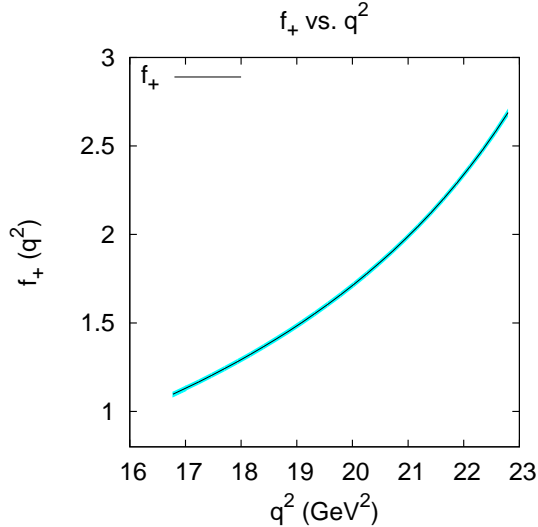}
\label{fig:fplus}
\caption{Continuum extrapolated $f_+$ for $B\to Kl^+l^-$ decay.  The width of the 
band indicates the statistical error; in addition, we estimate that the systematic 
errors will be $\sim 5$\%, depending somewhat on $q^2$.}
\end{figure}

After we obtain the continuum extrapolated $f_\parallel$  and $f_\perp$, 
we can construct $f_+$, which is crucial to the 
$B\to Kl^+l^-$ differential decay rate. The plot of $f_+$ is given in 
Fig.~\ref{fig:fplus}. 
Our simulation corresponds to the momentum transfer $q^2$ from 16${\rm GeV}^2$ to 23${\rm GeV}^2$. 
More study of the $q^2$ dependence of $f_+$ in the full kinematic range using 
the $z$-expansion  will be considered later~\cite{Boyd:1994tt}.

\section{Summary and Future Plans}
We report preliminary results from the study of the $B \to Kl^+l^-$ semileptonic 
decay process with statistical error only.
Current chiral-continuum extrapolations are done with NLO $SU(2)$ S$\chi$PT. 
More comparison of the $SU(3)$ 
and $SU(2)$ S\chipt will be studied. 
We will work on the systematic errors in the next step. These systematic errors come 
from the uncertainties in $r_1$ (used to set the scale) and physical 
quark masses, finite volume effect, chiral fits and so on. 
Finally, the $B\to Kl^+l^-$ process contains an additional form factor 
from the tensor current, which we will analyze in the future.

We thank Claude Bernard and Jack Laiho for helpful discussions on S$\chi$PT.
We also thank James Simone for the help on the renormalization constants. 
Computations for this work were carried out with resources provided by
the USQCD Collaboration, the Argonne Leadership Computing Facility,
the National Energy Research Scientific Computing Center, and the
Los Alamos National Laboratory, which are funded by the Office of Science of the
U.S. Department of Energy; and with resources provided by the National Institute
for Computational Science, the Pittsburgh Supercomputer Center, the San Diego
Supercomputer Center, and the Texas Advanced Computing Center, which are funded
through the National Science Foundation's Teragrid/XSEDE Program.
This work was supported in part by the U.S. Department of Energy under Grants
No.~DE-FG02-91ER40661 (S.G., R.Z.),
No.~DE-FG02-91ER40677 (R.D.J., A.X.K.). 
This manuscript has been co-authored by employees of Brookhaven Science
Associates, LLC, under Contract No. DE-AC02-98CH10886 with the U.S. Department of Energy.
R.S.V. acknowledges support from BNL via the Goldhaber Distinguished Fellowship.
Fermilab is operated by Fermi Research Alliance, LLC, under Contract
No.~DE-AC02-07CH11359 with the United States Department of Energy.

\end{document}